# Behavior of Lithium Oxide at Superionic Transition: First Principles and Molecular Dynamics Studies

M. K. Gupta, Prabhatasree Goel, R. Mittal, N. Choudhury and S. L. Chaplot

*Solid State Physics Division, Bhabha Atomic Research Centre, Trombay, Mumbai 400 085*

We report studies on the vibrational and elastic behavior of lithium oxide, $Li_2O$ around its superionic transition temperature. Phonon frequencies calculated using the *ab-initio* and empirical potential model are in excellent agreement with the reported experimental data. Further, volume dependence of phonon dispersion relation has been calculated, which indicates softening of zone boundary transverse acoustic phonon mode along [110] at volume corresponding to the superionic transition in $Li_2O$. The instability of phonon mode could be a precursor leading to the dynamical disorder of the lithium sub lattice. Empirical potential model calculations have been carried out to deduce the probable direction of lithium diffusion by constructing a super cell consisting of 12000 atoms. The barrier energy for lithium ion diffusion from one lattice site to another at ambient and elevated temperature has been computed. Barrier energy considerations along various symmetry directions indicate that [001] is the most favourable direction for lithium diffusion in the fast ion phase. This result corroborates our observation of dynamical instability in the transverse mode along (110) wave vector. Using molecular dynamics simulations we have studied the temperature variation of elastic constants, which are important to the high-temperature stability of lithium oxide.

I. INTRODUCTION

Lithium oxide exhibits high ionic conductivity whilst in solid condition above 1200 K, and belongs to the class of superionics, which allow macroscopic movement of ions through their crystal structure. This behavior is characterized by the rapid diffusion of a significant fraction of one of the constituent species within an essentially rigid framework formed by the other species. In lithium oxide, Li ion is the diffusing species and the current carrier, while oxygen ions constitute the rigid framework. In superionic conductors the ionic conductivity is greater than 0.01 (Ohm-cm)$^{-1}$. Lithium oxide, like other superionics, finds several technological applications[1,2]. These applications range from miniature lightweight high power density lithium ion batteries for heart pacemakers, mobile phones, laptop computers, *etc* to high capacity energy storage devices for next generation 'clean' electric vehicles. It is also a leading contender for future fusion reactors to convert energetic neutrons to usable heat and to breed tritium necessary to sustain D-T reaction[3]. This application is attributed to its high melting point, relatively low volatility and high Li atom density. Lithium oxide crystallizes in anti- fluorite, face centered cubic structure at ambient condition belonging to the space group $O^5_h$ (Fm3m)[4,5,6,7]. Lithium atoms are in tetrahedral sites. Its small size and low mass compared to oxygen permits to diffuse easily in the material at high temperature.

The novel high temperature properties of lithium oxide and similar compounds have warranted various experimental and first principle studies and other theoretical model calculations extensively[8-31]. An objective of these studies has been to understand the process of fast ion conduction and the role of defects in conductivity. Further interest has been to study Li diffusion from the point of view of tritium generation for future fusion reactors. The fact that this shows fast ion conduction while in solid state makes it an ideal link between solids and liquids.

Fluorites like $CaF_2$, $BaF_2$, $SrCl_2$, $PbF_2$ etc., show type II superionic transition. They attain high levels of ionic conductivity following a gradual and continuous disordering process within the same phase. In general, this extensive diffusion is assumed to be associated with the development of extensive Frenkel disorder in the anion sub lattice (cation in the case of anti-fluorites). This is characterized by a large decrease in the elastic constant[32,33] $C_{11}$ and a specific heat anomaly (a Schottky hump) at the transition temperature, $T_c$. Similar to the other fluorites, lithium oxide too shows a sudden decrease in the value of the $C_{11}$ elastic constant at the transition temperature, $T_c$ ~ 1200 K (the melting point of $Li_2O$ is 1705 K) although there does not seem to be any drastic change in the specific heat. The diffusion coefficient of lithium at this transition becomes comparable to that of liquids[34,35].

Extending on our earlier work[2] on this oxide using potential-based lattice and molecular dynamics, we have now performed *ab initio* phonon calculation. The main purposes of these calculations are to study the mechanism of diffusion of lithium in the superionic phase and comparative analysis of *ab-initio* and potential model calculation. The process of diffusion is related to the density of defects in the material, which increases with temperature. The decrease in the elastic constants has been studied using molecular dynamics simulations with increasing temperature. Figuring out diffusion pathways is crucial in understanding diffusion mechanism in a system. In our earlier studies we had inferred that Li moves from one tetrahedral site to another. In this work we have tried to deduce the most probable direction of diffusion by studying the barrier energy for Li movement along various high symmetry directions.

## II. COMPUTAIONAL DETAILS

The lattice dynamics calculations have been performed using both empirical potential as well as ab-initio methods. Empirical potential used is the same as our earlier work[2]. Density functional theory (DFT) has been shown to describe the structural and lattice dynamical properties of material using pseudopotentials and plane wave basis sets. In the frame work of density functional perturbation theory (DFPT) it is possible to calculate phonon frequencies, dielectric constants, piezoelectric and other properties. The calculations have been performed using Quantum espresso[36] package. We have chosen pseudopotential generated by using Perdew-Zunger exchange correlation functional under local-density approximation (LDA). It is a norm conserving pseudopotential. The energy cutoff for maximum plane wave energy has been chosen to be 120 Ry. The Brillouin zone integration has been performed using 12×12×12 K mesh. The k point mesh has been generated using Monkhorst-pack method.

By minimizing the total energy with respect to lattice parameter we have obtained the equilibrium lattice constant. The equilibrium lattice constant comes out to be 4.45 Å for LDA at 0 K and ambient pressure while the experimental lattice parameter[31] is 4.60 Å at 300 K. LDA exchange correlation functional underestimate the lattice constant by 2-3%. The difference between the calculated and experimental values is ~ 3% which is acceptable under LDA. The phonon calculation has been done under the constraint of crystal acoustic sum rule. The slopes of the acoustic branches along (100) and (110) are used for the calculation of elastic constants and bulk modulus.

The calculation of thermal expansion coefficient has been carried out in the quasi harmonic approximation (QHA). In QHA thermal expansion coefficient $\alpha_V$ is related to the Grüneisen parameter by the following relation:

$$\alpha_V(T) = \frac{1}{BV}\sum_i \Gamma_i C_{V_i}(T) \qquad \Gamma_i = -\frac{\partial(\ln \omega_i)}{\partial(\ln V)}$$

Where V is the volume of the unit cell, B is the bulk modulus, $\Gamma_i$ is the Grüneisen parameter due to $i^{th}$ phonon mode and $C_{V_i}(T)$ is the specific heat due to $i^{th}$ phonon mode at temperature T. Volume dependence of phonon frequencies has been calculated at equilibrium unit cell volume and increasing and decreasing the volume by 0.5%.

The temperature variation of the elastic properties has been calculated using the molecular dynamics (MD) technique. These calculations are carried out using a super cell (8 × 8 × 8) consisting of 6144 atoms. The interatomic potential is the same as used in our earlier work[2].

### III. RESULTS AND DISCUSSION

**A. Phonon Dispersion Relation**

The phonon dispersion using the potential model at lattice parameters corresponding to ambient (a = 4.66 Å, 300 K) and superionic (a = 4.90 Å, 1600 K) temperatures are shown in Fig.1. These values of lattice parameters have been obtained from MD simulations. As expected the phonon frequencies along all the three direction are found to soften with increase of temperature. The softening is found to be small for all the modes except for the lowest transverse acoustic branch along [110] at zone boundary. This branch shows large softening and the frequency of zone boundary mode along (110) is imaginary at this volume.

The calculated phonon dispersion relation corresponding to the experimental lattice parameter a = 4.6 Å (at 300 K) using *ab-initio* technique is also shown in Fig. 2. The results are in good agreement with the classical lattice dynamics calculations and reported experimental data. The compound exhibits superionic transition in the vicinity of 1200 K. Hence we have performed *ab-initio* phonon calculations at various unit cell parameters. Here again phonon modes soften on increase of lattice parameter. We find that zone boundary transverse acoustic (TA) mode along [110] direction become imaginary corresponding to the lattice parameter of a= 4.88 Å. We note that the results obtained from the potentials are fully consistent with ab initio results. The potentials could therefore be treated as derived from first principles.

The eigenvector of TA mode has been plotted (Fig. 3) corresponding to the unit cell parameter of a= 4.88 Å. We find that lithium atoms in the alternate layers move opposite to each other along [001] while oxygen's are at rest. The energy barrier for diffusion along [001] has minimum value (as discussed below). Hence increasing the temperature could lead to migration of lithium ions from one site to another vacant site along [001] direction, which can easily be visualized from Fig. 3. Beyond 1200 K, some of the lithiums might just have sufficient energy to move from their ideal positions and start diffusing[2]. It is possible that the softening of these modes might be the precursor to the process of diffusion. Fracchia[13] et al have also reported softening of zone boundary mode along [001] in $Li_2O$.

Fig. 4 gives the change in the transverse acoustic frequency with increasing lattice parameter, as calculated from both the *ab-initio* and potential model calculations. As discussed above the lowest transverse acoustic mode along [110] at zone boundary is found to soften sharply at volume corresponding to 1200 K**.** Incidentally this is the region of the supersonic diffusion in lithium dioxide.

**B. Temperature Variation of Elastic Constants and Thermal Expansion**

The elastic constants obtained using first principles approach have been compared with those obtained from potential model and reported experimental results[31,32] in Table I. Bulk modulus has been calculated using the relation B= $(C_{11}+2C_{12})/3$, where $C_{11}$ and $C_{12}$ are the elastic constants. The calculated elastic constants $C_{11}$, $C_{12}$, $C_{44}$ and bulk modulus B using *ab-initio* method at 0 K are in good agreement with reported experimental data at 300 K and our previous[2] lattice dynamics calculations.

The process of diffusion is related to the density of defects in the material. It is well known that Frenkel disorder increases with temperature. Molecular dynamics (MD) technique is used to calculate the effect of temperature on the elastic properties. Calculation of time correlation functions gives dynamic information from molecular dynamics studies. For the study of individual phonons, Fourier components of atomic trajectories have to be considered. We have calculated dynamical structure factor $S(\mathbf{Q},\omega)$ along several propagation directions. The structure factor is related to the time correlation function $F(\mathbf{Q},t)$. We find that increasing temperature results in significant changes in the acoustic phonon frequencies which in turn bring a change in the elastic properties. The calculations are found to be in good comparison (Fig. 5) with the reported experimental data. $C_{11}$ shows maximum softening which is one of the characteristic features of a superionic compound. The change in $C_{12}$ is marginal while $C_{44}$ remains almost constant with temperature increase. The gradual change in $C_{44}$ and $C_{12}$ with increasing temperature is brought out better in the molecular dynamics simulation and is in good agreement with experimental findings. The details of dependence of elastic constants on temperature play a pivotal role in understanding the role and uses of lithium oxide as blanket material for tritium production in fusion reactors.

The thermal expansion coefficient of lithium oxide from the ab-initio quasi-harmonic lattice dynamics calculation is shown in Fig. 6. These results are in excellent agreement with previous potential model calculations. We find that at low temperatures up to 600 K, *ab-initio* as well as potential model quasiharmonic dynamical calculations are in agreement with experimental data. However, at high temperature above 600 K there is a deviation between the thermal expansion obtained from the experiments and the potential model phonon calculations. This may be due to the fact that at high temperature explicit anharmonicity of phonons play an important role. The effect has not been included in the thermal expansion calculated from the volume dependence of phonon energies (implicit contribution). This is the reason that calculated thermal expansion from MD is in better agreement with the experimental data at high temperatures.

**C. Barrier Energy for Li Diffusion**

The energy needed for Li migration from one site to its neighbouring site is the barrier energy. This movement or diffusion depends on the interatomic bonding, temperature, microstructure, size and type of atoms diffusing. We have calculated the energy required to move lithium ion from one lattice site to another along different high symmetry directions, namely [001], [110] and [111] using a 10×10×10 supercell. Barrier energy values have been obtained by creating a vacancy and moving a lithium atom from its initial lattice site towards the vacant lattice site along a given high symmetry direction, and monitoring the change in the total crystal potential. This enables us to identify the preferred direction of diffusion and probability of diffusion at different temperatures. The charge neutrality of the super cell has been maintained in the complete process. The profiles of the energy changes are shown in fig. 7. The calculated barrier energies for movement of Li atom along [001], [110] and [111] directions are 0.2(0.0), 9.0(6.0) and 2.0(1.8) eV respectively at unit cell volume corresponding to 300(1600) K. It can be seen that the most favorable direction for lithium movement is along [001] direction, with a smaller possibility of migration along [111], and a negligible possibility along [110]. But this does not rule out lithium's motion along a combination of other directions. A direct movement of lithium from one lattice position to a neighboring vacant site is most likely to occur along [001] direction. Here, at ambient conditions, the barrier energy is about one order of magnitude lesser in comparison with other high symmetry direction values. It becomes negligible in the superionic phase.

Our studies on phonon dispersion around superionic region indicate an initial softening of transverse acoustic mode near zone boundary along [110] direction. This seems to be a precursor to diffusion of lithium as only lithium ions move in this mode while oxygens remain unmoved. Since [001] direction is orthogonal to [110], it is plausible that on acquiring energies sufficient to overcome the barrier energy along this direction, lithium starts to migrate from one lattice site to another. Although [001] appears to be the most preferred direction, it is possible that lithium may migrate along other directions too albeit with lesser probability. The increase in defects with increasing temperature further helps in the formation of vacancies which enhance diffusion further.

## IV. CONCLUSIONS

A combination of empirical and first principle studies have been used to successfully explain the behavior of lithium oxide at the onset of superionic transition. Computed phonon dispersion along symmetry directions and thermal expansion are in good agreement with reported experimental data. We find that around fast ion transition temperature, zone boundary TA phonon mode along [110] becomes unstable. This dynamical instability could be a precursor to the diffusion of lithium ions as suggested by the eigenvector of the soft phonon mode. We also are able to bring out the decrease in elastic constants with increasing temperature, which is one of the characteristic features exhibited by these fast ion conductors. Finally we have calculated the decrease in the barrier energy with temperature along [001] direction.

TABLE I. Comparison of elastic and structural data obtained using ab-initio approach with available experimental[31,32] results and potential model[2] calculations.

| Parameters | Expt. | Ab-initio | Potential model |
|---|---|---|---|
| Lattice constant (Å) | 4.606 | 4.45 | 4.61 |
| $C_{11}$ (GPa) | 202 | 234 | 213 |
| $C_{44}$ (GPa) | 59 | 72 | 52 |
| $C_{12}$ (GPa) | 21.5 | 31.5 | 56 |
| B (GPa) | 82 | 99 | 103 |

FIG. 1. (Color online) Phonon dispersion using interatomic potential in the quasi-harmonic approximation. The calculations have been carried out corresponding to experimental lattice parameters at 300 K (4.66 Å) and 1600 K (4.88 Å). These values of lattice parameters have been obtained from molecular dynamics simulations. Phonon energies shown below E = 0 have imaginary values indicating unstable modes.

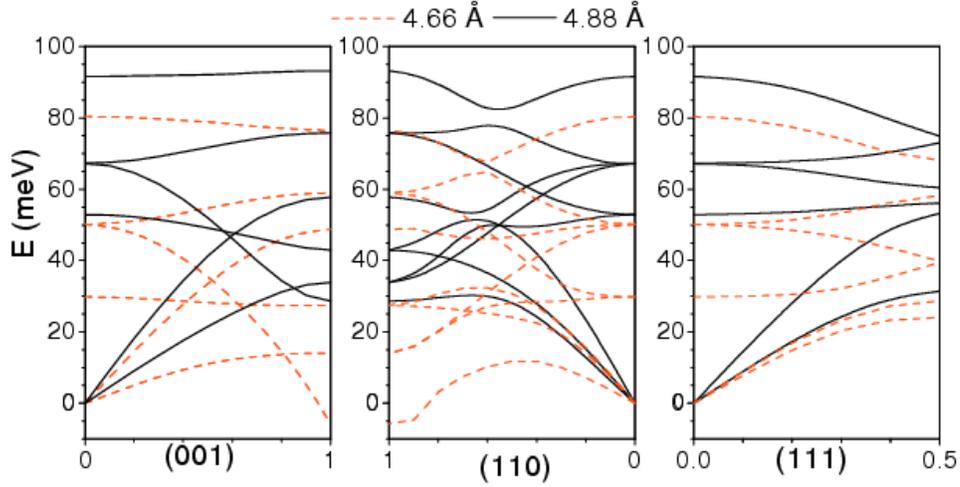

FIG. 2. (Color online) Phonon dispersion using ab-initio in the quasi-harmonic approximation. The full and dashed lines correspond to calculations performed at experimental a= 4.6 Å (0 K) and a= 4.88 Å (1600 K). The solid symbols correspond to reported experimental[32] data. The lowest acoustic branch along (001) and (110) both become imaginary around the fast ion transition region. Phonon energies shown below E = 0 have imaginary values indicating unstable modes.

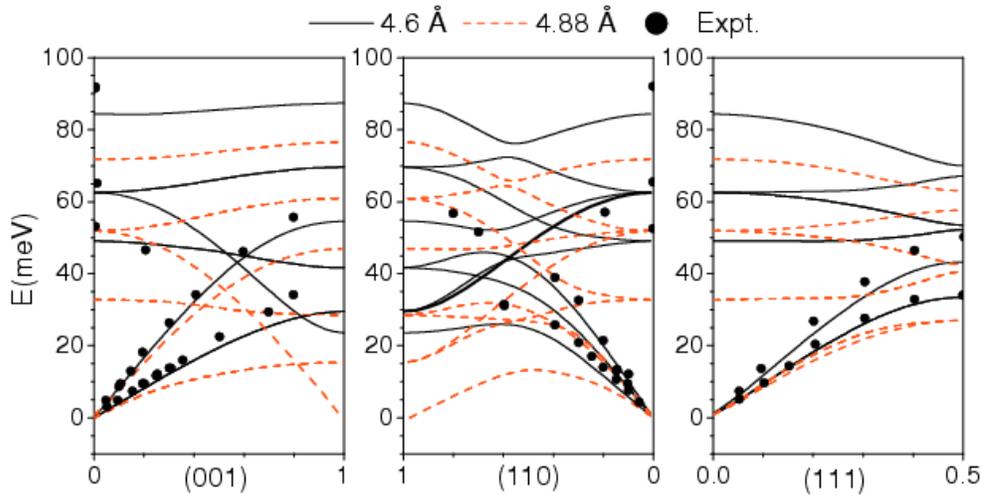

FIG. 3. (Color online) Motion of individual atoms for zone boundary mode along [110] direction at lattice parameter corresponding to a= 4.88 Å. The lengths of arrows are related to the displacements of the atoms. The absence of an arrow on an atom indicates that the atom is at rest. b-axis is perpendicular to the plane. Key; O: red spheres, Li: blue spheres.

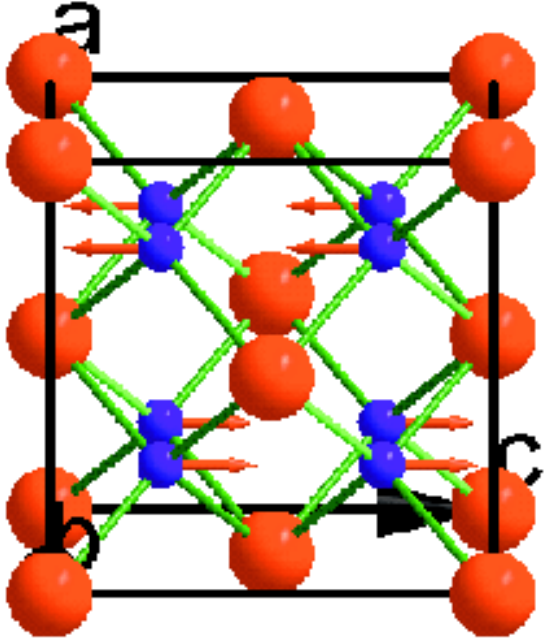

FIG. 4. (Color online) Softening of zone boundary transverse acoustic (TA) phonon along [110]. Phonon energies shown below E = 0 have imaginary values indicating unstable modes.

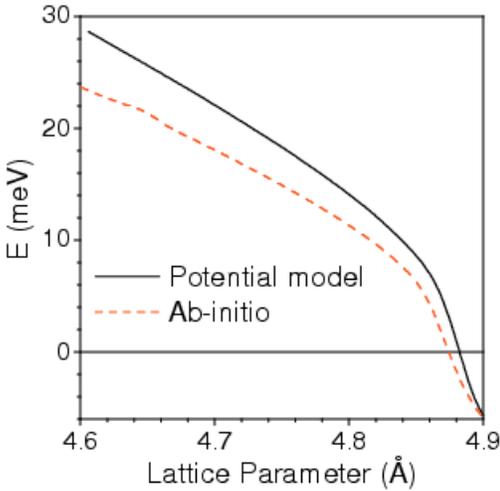

FIG. 5. (Color online) Softening of elastic constants with increasing temperature compared with reported experimental[32] results. The open, half filled symbols correspond to the calculated values using molecular dynamics (MD) and lattice dynamics (LD) formalism while solid symbols denote the reported experimental results.

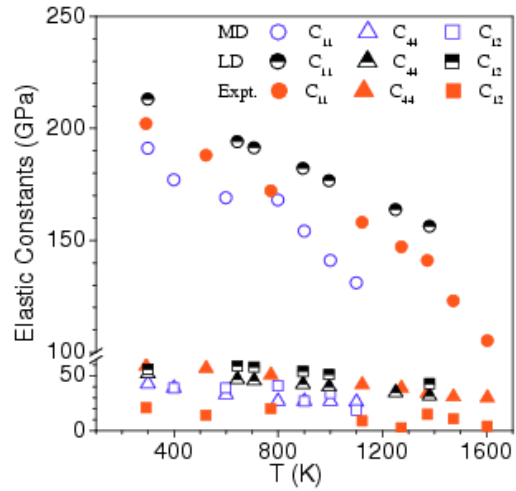

FIG. 6. (Color online) Comparison between the experimental[1] and calculated volume thermal expansion of lithium oxide, $Li_2O$. The dotted and full lines correspond to quasi-harmonic lattice dynamics results from *ab initio* and potential model calculations respectively; open and solid symbols denote calculated molecular dynamics[2] and reported experimental data.

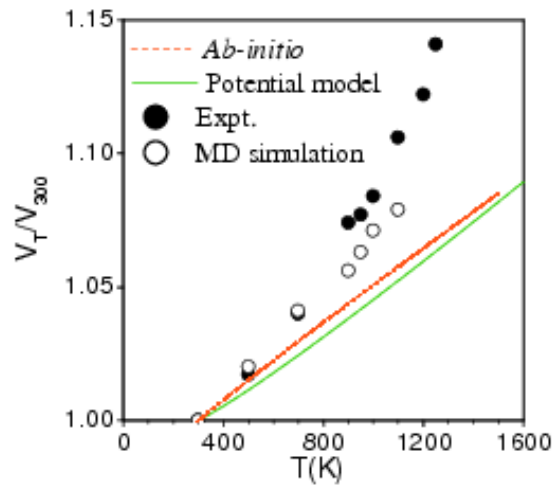

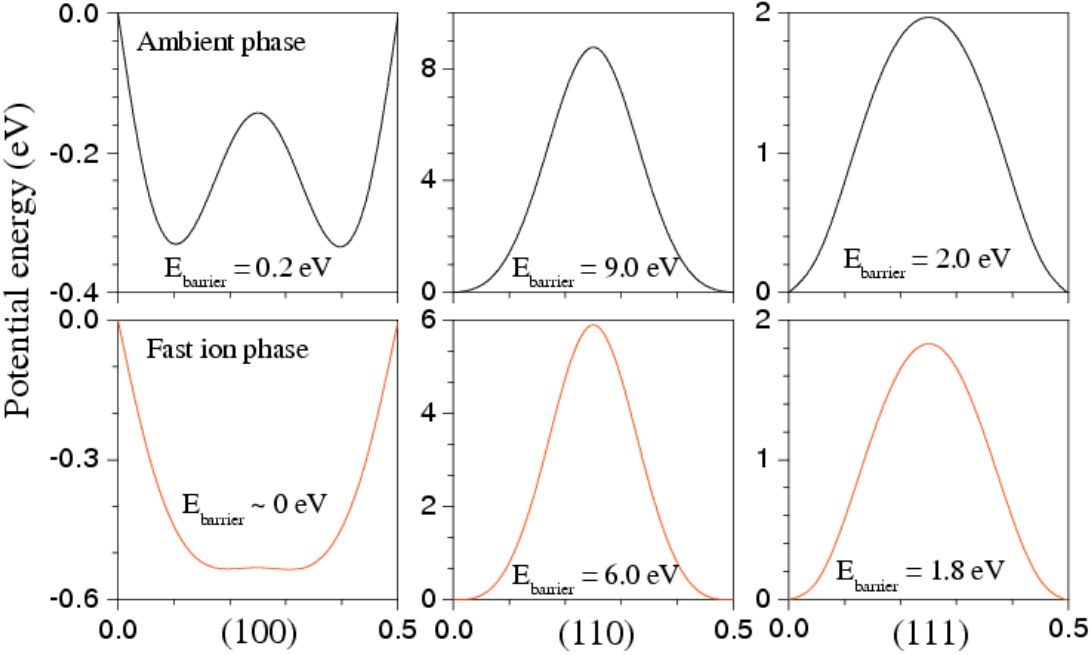

FIG. 7. (Color online) Potential energy barriers for lithium movement along high symmetry directions at ambient and around superionic temperatures. Calculations have been carried out using 10×10×10 super cell corresponding to lattice parameter values of 46.06 Å and 48.8 Å.